\documentclass[10pt,a4paper]{article}
\usepackage[utf8]{inputenc}
\usepackage[english]{babel}
\usepackage{graphicx}
\usepackage{hyperref} 
\usepackage{geometry}
\usepackage{amsmath}
\usepackage{amssymb}
\usepackage{amsfonts}
\usepackage{multicol}
\usepackage{float}
\usepackage{caption}
\usepackage{subcaption}
\usepackage{cite}
\numberwithin{equation}{section}
\usepackage{amsthm}
\usepackage[utf8]{inputenc}
\usepackage{xcolor}
\newcommand{\gn}{G_{\rm N}}

\newcommand{\lp}{l_{\rm p}}

\title{\bf Singularities in Quantum Corrected Space-times}
\author{Xavier Calmet$^{a}$\thanks{E-mail: X.Calmet@sussex.ac.uk},$\ $
Roberto Casadio$^{b,c}$\thanks{E-mail: Roberto.Casadio@bo.infn.it}$\ $ 
and Folkert~Kuipers$^{a,b}$\thanks{E-mail: F.Kuipers@sussex.ac.uk} 
\\
\\
{\em$^a$ Department of Physics and Astronomy, University of Sussex,}\\ 
{\em Brighton, BN1 9QH, United Kingdom}
\\
\\
{\em $^b$ Dipartimento di Fisica e Astronomia, Alma Mater Universit\`a di Bologna,}\\
{\em via Irnerio 46, 40126 Bologna, Italy}
\\
\\
{\em $^c$ I.N.F.N., Sezione di Bologna, I.S.~FLAG}\\
{\em viale B.~Pichat~6/2, 40127 Bologna, Italy}
}
\begin{document}
\maketitle
%
\begin{abstract}
In this paper we consider a static and regular fluid generating a locally spherically symmetric and time-independent space-time and calculate the leading quantum corrections to the metric to first order in curvature.
Starting from a singularity free classical solution of general relativity, we show that  singularities can be introduced in the curvature invariants by quantum gravitational corrections calculated using an effective field theory approach to quantum gravity. We identify non-trivial conditions that ensure that curvature invariants remain singularity free to leading order in the curvature expansion of the effective action.
\end{abstract}
\flushbottom
%
\thispagestyle{empty}
\pagebreak
\pagenumbering{arabic}
\section{Introduction}
Black holes are stationary vacuum solutions of Einstein's field equations.
Despite the simplicity of these solutions, rotating black holes which are described by the Kerr metric
appear to account for the numerous observations of astrophysical black holes with a remarkable accuracy.
Furthermore, it is known that massive stars collapse at the end of their lifetime and form black holes.
The idea that gravitational collapse leads to a black hole is strengthened by singularity theorems
that prove geodesic incompleteness when a trapped surface is formed~\cite{Penrose:1964wq, Hawking:1966vg}. 
However, the collapse picture may not be fully realistic. Solutions like the Kerr metric are in fact vacuum solutions, and delta-like sources are not well defined in general relativity~\cite{Geroch:1987qn}. The collapse of a star to a Kerr black hole thus requires the destruction of the matter that made up the star, while some information like the mass and angular momentum is conserved. Indeed this is the reason why general relativity is considered to break down at singularities.
\par
It is usually assumed that a theory of quantum gravity will provide a physical mechanism to resolve these classical curvature singularities.
Although many candidate theories of quantum gravity have been developed, an ultra-violet complete theory
of quantum gravity is still illusive.
There exists however a unique infra-red theory of quantum
gravity~\cite{Weinberg:1980gg, Barvinsky:1984jd, Barvinsky:1985an, Barvinsky:1987uw, Barvinsky:1990up,
Buchbinder:1992rb, Donoghue:1994dn}
that is valid up to the scale where the new physics necessary for an ultra-violet completion kicks in.
This scale is known to be far beyond the reach of current experiments and is assumed to be at the Planck scale
unless there is a very large number of fields in the model.
\par
An important prediction of the effective field theory of quantum gravity is the leading quantum correction to
the Newtonian potential~\cite{Stelle:1977ry,Donoghue:1993eb,Calmet:2018qwg}.
This correction can equivalently be described by the introduction of two classical fields.
Recently it has been shown that these fields can lead to the violation of assumptions of the singularity theorems
and that singularities can therefore be avoided before Planck scale energies are
reached~\cite{Kuipers:2019qby, Kuntz:2019lzq}.
Moreover the possibility of the singularity avoidance in a gravitational collapse \cite{Frolov:1981qr}
and a hypothetical big crunch~\cite{Donoghue:2014yha} was shown earlier in the same framework.
\par
It is thus possible that a space-time that is classically singular becomes regular, when perturbative
quantum gravitational effects are taken into account.
In the specific cases studied in Refs.~\cite{Frolov:1981qr,Donoghue:2014yha}, singularities avoidance happens
at energy densities that do not exceed the Planck scale and can thus be described by the effective action approach
to quantum gravity.
\par
In this paper we investigate the opposite scenario in which curvature singularities can be introduced
on regular space-times when the quantum gravitational corrections are taken into account. Since we are working in an effective field theory framework, this question should be rephrased as: can quantum corrections to the curvature invariants reach the Planck scale, if the classical curvature is well behaved, i.e. singularity free? It is not possible within the effective field theory approach to draw strong conclusions about the fate of such new singularities, as the logic of any perturbative approach dictates to dismiss the results in regimes where the perturbation theory is no longer under control. Nevertheless this is an intriguing question, as non-perturbative quantum gravity is not yet well-understood. Moreover, from an effective field theory perspective singularities arising at any order in perturbation theory should be treated at the same level as the classical singularities. Indeed in the effective field theory framework general relativity is the zeroth order approximation of a theory of quantum gravity, and any higher order theory is considered to be an improvement over this low energy approximation.
\par
In this work we derive non-trivial conditions for which classical regular space-times remain regular in a first order effective field theory approach. Furthermore, we provide an explicit example of a space-time that is classically regular but contains a singularity at first order.
\par
This paper is organized as follows:
in the next section we introduce the general form for the metric considered in this paper;
in section~\ref{s:ec} we discuss some properties of this metric, related to the pressure and density
of its matter source.
These properties can be used to put further constraints on the metric;
section~\ref{s:qc} discusses the leading quantum corrections to this metric and derives conditions
for which a classical regular metric remains regular, if the leading quantum corrections are taken into account;
in section~\ref{s:conc} we conclude and appendix~\ref{A} discusses the Bardeen metric as an explicit example.
\section{A general metric}
\label{s:metric}
We consider a Lorentzian $(3+1)$-dimensional space-time containing a regular fluid.
We choose the origin of our local coordinate system at a local maximum in the density of the fluid
and assume the space-time to be locally spherically symmetric and time-independent around the origin.
The line element can then be written as~\footnote{Since we assume only local spherical symmetry and
time-independence, this line element needs only be valid in a small region around $r=0$.}
\begin{equation}
\label{eq:metric}
	ds^2 = - f(r) \,dt^2 + \frac{dr^2}{g(r)}  + r^2 \left(d\theta^2 + \sin\theta^2\, d\phi^2
\right)	
\ ,
\end{equation}
in which we employ the usual areal radial coordinate $r$.
Consistently with the regular matter source, we assume the space-time to be smooth and
regular everywhere, which we define in this paper by $|R|$,
$|R_{\mu\nu}\,R^{\mu\nu}|$, $|R_{\mu\nu\rho\sigma}\,R^{\mu\nu\rho\sigma}|<\infty$.~\footnote{Of course,
$R_{\mu\nu\rho\sigma}$ is the Riemann tensor, $R_{\mu\nu}$ the Ricci tensor and $R$ the Ricci scalar.
The above conditions immediately imply for the Weyl tensor $|C_{\mu\nu\rho\sigma}\,C^{\mu\nu\rho\sigma}|<\infty$.}
If we impose these conditions, we can expand $f$ and $g$ around $r=0$ in the following way:
\begin{align}        
	f(r) &= 1+ \sum_{n=0}^{\infty} a_{2n}\, r^{2n} + f_\infty(r)
	\ ,
	\label{eq:smoothmetricf}
	\\                  
	g(r) &= 1+ \sum_{n=1}^{\infty} b_{2n}\, r^{2n} + g_\infty(r)
	\ ,
	\label{eq:smoothmetricg}
\end{align}
where the Lorentzian signature requires $a_0>-1$.
Furthermore, the non-analytic parts $f_\infty$ and $g_\infty$ have the property
\begin{equation}
	\lim_{r\rightarrow 0} \frac{f_\infty(r)}{r^n} = \lim_{r\rightarrow 0} \frac{g_\infty(r)}{r^n} = 0
	\ , 
	\qquad \forall\; n\in \mathbb{N}.
\end{equation}
and
\begin{equation}
	\exists\; \epsilon>0
	\quad {\rm s.t.} \quad
	 f_\infty(r)=f_\infty(-r)
	\quad
	{\rm and}
	\quad
	g_\infty(r)=g_\infty(-r)
	\ ,
	\quad 
	\forall \; r\in[0,\epsilon)
\ .
\end{equation}
Since we only want to perform a local analysis, we can truncate the series so that
\begin{align}                                             
	f(r) &= 1 + a_0 + a_{m}\, r^{m} + \mathcal{O}\left(r^{m+2}\right)
	\ ,
	\\                                                 
	g(r) &= 1 + b_{n}\, r^{n} + \mathcal{O}\left(r^{n+2}\right)
	\ ,
\end{align}
where $m,n\geq2$, $a_0>-1$, $a_m,b_n\neq0$.
Furthermore, we say that $m=\infty$, $n=\infty$, if $f(r)= 1 + a_0 + f_\infty(r)$,
$g(r)= 1 + g_\infty(r)$ respectively.
\section{Energy conditions}
\label{s:ec}
For the line element introduced in the previous section, the regular
energy density, radial and transversal pressure
 generating the space-time metric~\eqref{eq:metric} with $f$ and $g$ given 
in Eqs.~\eqref{eq:smoothmetricf} and \eqref{eq:smoothmetricg}, respectively,
can be calculated from the Einstein tensor and read
\begin{align}
	\rho 
	&= - \frac{1}{8\,\pi \,\gn}\, b_n\, (n+1)\, r^{n-2} + \mathcal{O}(r^{n-1})
	\ ,
	\\
	p_{\parallel} 
	&= \frac{1}{8\,\pi\, \gn} \left(
	\frac{a_m}{1+a_0} \, m \, r^{m-2} 
	+ b_n \, r^{n-2} \right) 
	+ \mathcal{O}(r^{m-1},r^{n-1})
	\ ,
	\\
	p_{\perp} 
	&= \frac{1}{8\,\pi\, \gn} \left(
	\frac{a_m}{1+a_0} \frac{m^2}{2} \, r^{m-2} 
	+ b_n \,\frac{n}{2}\, r^{n-2} \right) 
	+ \mathcal{O}(r^{m-1},r^{n-1})
	\ .
\end{align}
A positive energy density thus requires $b_n<0$.
Moreover, a non-zero energy density and pressure at $r=0$ requires $n=2$.
Furthermore, $|p_\perp|\geq|p_\parallel|$, and a necessary but not sufficient requirement for equality
(hence isotropy) is that one of the following is true:
\begin{itemize}
	\item $\left(m=2 \vee m=\infty \right)\; \wedge \; \left(n=2 \vee n =\infty\right)\ $,
	\item $m=n \; \wedge \; a_m \, m + b_n \, (1+a_0) = 0\ $.
\end{itemize}
We can write,
\begin{align}
	p_\parallel &= - \frac{\rho}{n+1} \left(1 + \frac{a_m\, m}{b_n\,(1+a_0)}\, r^{m-n} \right) 
	+ \mathcal{O}(r^{m-n-1},r^{-1})\\
	&\equiv w\, \rho.
\end{align}
Depending on the parameter of the model one can find various values for $w$. 
It follows that a de~Sitter core with $w=-1$ requires
\begin{equation}
	m=n \; \wedge \; \frac{a_m}{b_n\,(1+a_0)} = 1
	\ ,
	\label{eq:dscore}
\end{equation}
a non-relativistic matter (dust) core with $w=0$ requires
\begin{equation}
	m=n \; \wedge \; \frac{a_m}{b_n\,(1+a_0)} = - \frac{1}{n}
	\ ,
\end{equation}
and an ultra-relativistic (radiation) core with $w=\frac{1}{3}$ requires
\begin{equation}
	m=n \; \wedge \; \frac{a_m}{b_n\,(1+a_0)} = -\frac{n+4}{3\,n}.
\end{equation}
Finally an asymptotically Minkowski core with $\rho=0$ and $|w|<\infty$ requires $m\geq n>2$.
\section{Quantum corrections to the metric}
\label{s:qc}
We shall here use the same approach as discussed in Refs.~\cite{Calmet:2017qqa,Calmet:2019eof},
for which we review the main steps.~\footnote{Note that the signature conventions in this work
differ from those in~\cite{Calmet:2017qqa,Calmet:2019eof}.}
The effective action is given by
\begin{equation}
	\Gamma = \Gamma_{\rm L}[g] 
	- \Gamma_{\rm NL}[g]
	+ S_{\rm M}
	+\mathcal{O}(\gn)
\end{equation}
with $S_{\rm M}$ the matter action and
\begin{align}
	\Gamma_{\rm L}[g]
	&= \int d^4 x\, \sqrt{g} \left[ 
	\frac{R}{16\,\pi\, \gn}
	+ \tilde{c}_1(\mu)\, R^2
	+ \tilde{c}_2(\mu)\, R_{\mu\nu}\, R^{\mu\nu} \right]
	\ ,
	\\
	\Gamma_{\rm NL}[g]
	&= \int d^4 x\, \sqrt{g}
	\left[
	\tilde{\alpha}\, R \ln\!\left(-\frac{\Box}{\mu^2}\right) R
	+ \tilde{\beta}\, R_{\mu\nu} \ln\left(-\frac{\Box}{\mu^2}\right) R^{\mu\nu} \right]
	\ ,
\end{align}
where $\tilde{c}_i(\mu)$ are renormalization scale dependent coefficients that follow from matching
with an ultra-violet complete theory and experiment.
Furthermore, $\tilde{\alpha} = \alpha-\gamma$, $\tilde{\beta}= \beta + 4\,\gamma$ with the values of $\alpha$,
$\beta$, $\gamma$ given in Table~\ref{coeff1}.
The equation of motion for the metric can then be solved perturbatively in $\lp^2=\hbar\,\gn$
(and we will set $\hbar=1$ when no ambiguity can arise).
The zeroth order equation is the Einstein equation
\begin{table}
	\center
	\begin{tabular}{| c | c | c | c |}
		\hline
		& $ \alpha $ & $\beta$ & $\gamma$  \\
		\hline
		\text{Scalar} & $ 5(6\xi-1)^2$ & $-2 $ & $2$     \\
		\hline
		\text{Fermion} & $-5$ & $8$ & $7 $ \\
		\hline
		\text{Vector} & $-50$ & $176$ & $-26$ \\
		\hline
		\text{Graviton} & $250$ & $-244$ & $424$\\
		\hline
	\end{tabular}
	\caption{Non-local Wilson coefficients for different fields. 
All numbers should be divided by $11520\pi^2$.
Furthermore, $\xi$ denotes the value of the non-minimal coupling for a scalar theory.
The values  for the scalar, fermion and vector field have been calculated in Refs.~\cite{Deser:1974cz,Birrell:1982ix}.
The values for the graviton can be gauge dependent due to the graviton self-interaction diagrams~\cite{Kallosh:1978wt}.
However, it is possible to define a unique effective action with gauge independent coefficients leading
to the gauge independent results quoted here~\cite{Barvinsky:1984jd,Vilkovisky:1984st,Barvinsky:1985an}.}
	\label{coeff1}
\end{table}
\begin{equation}
	G_{\mu\nu} = 8\, \pi\, \gn\, T_{\mu\nu}
	\ ,
\label{eq:EinsteinOrder0}	
\end{equation}
and the first order equation is given by
\begin{equation}
\label{eq:EinsteinOrder1}
	G_{\mu\nu}^{\rm L}
	+ 16 \,\pi\, \gn \left( H_{\mu\nu}^{\rm L} - H_{\mu\nu}^{\rm NL} \right)
	= 0,
\end{equation}
where
\begin{align}
	G_{\mu\nu}^{\rm L}
	&= -\frac{1}{2} \left(	
	\Box g_{\mu\nu}^{\rm q} - g_{\mu\nu} \Box g^{\rm q}
	+ \nabla_\mu \nabla_\nu g^{\rm q} 
	+ 2 R^{\alpha~\beta}_{~\mu~\nu} g_{\alpha\beta}^{\rm q}
	\right.\nonumber\\
	&\qquad \qquad \qquad \left.
	- \nabla_\mu \nabla^\beta g_{\nu\beta}^{\rm q} 
	- \nabla_\nu \nabla^\beta g_{\mu\beta}^{\rm q}
	+ g_{\mu\nu} \nabla^\alpha \nabla^\beta g_{\alpha\beta}^{\rm q} \right)
	\ ,
	\\
	H_{\mu\nu}^{\rm L} 
	&= 2\,\tilde{c}_1 \left( 
	R\, R_{\mu\nu} 
	- \frac{1}{4} g_{\mu\nu} R^2 
	+ g_{\mu\nu} \Box R
	- \nabla_\mu \nabla_\nu R \right) \nonumber\\
	& \quad
	+ \tilde{c}_2 \left( 
	2 R_{~\mu}^\alpha R_{\nu\alpha} 
	- \frac{1}{2} g_{\mu\nu} R_{\alpha\beta} R^{\alpha\beta}
	+ \Box R_{\mu\nu} \right. \nonumber\\
	& \qquad \qquad \qquad \left.
	+ \frac{1}{2} g_{\mu\nu} \Box R
	- \nabla_\alpha \nabla_\mu R_{~\nu}^\alpha
	- \nabla_\alpha \nabla_\nu R_{~\mu}^\alpha \right)
	\ ,
	\\
	H_{\mu\nu}^{\rm NL} 
	&=  2 \tilde{\alpha} \left( 
	R_{\mu\nu} 
	- \frac{1}{4} g_{\mu\nu} R
	+ g_{\mu\nu} \Box
	- \nabla_\mu \nabla_\nu \right)
	\ln\left(-\frac{\Box}{\mu^2}\right) R
	\nonumber \\
	& \quad 
	+ \tilde{\beta} \left( 
	\delta_\mu^\alpha\, R_{\nu\beta}
	+ \delta_\nu^\alpha\, R_{\mu\beta}
	- \frac{1}{2} g_{\mu\nu} R_{~\beta}^\alpha
	+ \delta_{\mu}^\alpha g_{\nu\beta} \Box \right.\nonumber\\
	& \qquad \qquad \qquad
	+ g_{\mu\nu} \nabla^\alpha \nabla_\beta 
	- \delta_\mu^\alpha \nabla_\beta \nabla_\nu
	- \delta_\nu^\alpha \nabla_\beta \nabla_\mu \bigg)
	\ln\left(- \frac{\Box}{\mu^2}\right) R_{~\alpha}^\beta
	\ .
	\label{eq:NL}
\end{align}
Equation~\eqref{eq:EinsteinOrder1} can now be solved for the leading quantum corrections to the metric such that
\begin{equation}
	\tilde{g}_{\mu\nu} = g_{\mu\nu} + \hbar\, \gn\, \delta g_{\mu\nu}
	\ ,
\end{equation}
where $\tilde{g}_{\mu\nu}$ is the quantum corrected metric, $g_{\mu\nu}$ is the solution of Eq.~\eqref{eq:EinsteinOrder0}
and $\delta g_{\mu\nu} \equiv g_{\mu\nu}^q$.
\subsection{Local corrections}
The leading local quantum corrections are found to be
\begin{align}
	\delta g_{tt}^{\rm L} 
	=& - 32 \,\pi\, \gn \left[\tilde{c}_1\, a_m\, m\, (m+1)\, r^{m-2}
	+ (2\tilde{c}_1 + \tilde{c}_2)\, b_n\, (1+a_0)\, (n+1)\,  r^{n-2}\right]
	\nonumber
	\\
	&
	+ \mathcal{O}\left(r^{m-1},r^{n-1}\right)
	\ ,
	 \label{eq:LocCorTemp}
	\\
	\delta g_{rr}^{\rm L} 
        =&
        -16\, \pi\, \gn \left[(2\tilde{c}_1 + \tilde{c}_2)\, \frac{a_m}{1+a_0}\, m\, (m^2 - m - 2)\, r^{m-2}
        \right.
        \nonumber
        \\
	& \qquad \qquad \quad
	+ (4\tilde{c}_1 + \tilde{c}_2)\, b_n\, (n^2 - n - 2)\, r^{n-2} \bigg]
	+ \mathcal{O}\left(r^{m-1},r^{n-1}\right)
	\ .
	\label{eq:LocCorRad}
\end{align}
Hence, the leading local corrections are of the order $r^{m-2}$ and $r^{n-2}$.
For $m\leq3$ or $n\leq3$ these corrections would make the space-time singular at $r=0$.
Interestingly for the special cases $m,n=2$ we find an exact cancellation keeping the space-time regular.
Moreover, due to the assumptions of local spherical symmetry, regularity and smoothness of the classical
space-time, we imposed the conditions $m,n\geq2$ and $m,n$ even from the onset.
Therefore the local corrections do not pose any threats to the regularity and smoothness of the space-time
at this order.
\subsection{Non-local corrections}
Corrections due to the non-local terms in Eq.~\eqref{eq:NL} are more difficult to calculate, if the line element
is only known locally, since the $\ln\Box$ is an infinite derivative operator.
However, for a smooth time-independent and spherically symmetric function $f$, one can derive an expansion
in a flat background given by~\footnote{Since we assume a regular geometry, $\ln\Box$ can be expanded
as a power series in $\gn$.
The leading term in such an expansion is determined by the flat space kernel of $\ln\Box$.
Corrections due to curvature are subleading.}~\footnote{Calculation of such expressions has a long history
in both the mathematics and physics literature.
For the calculation of this particular expression we follow the same steps as presented in appendix~A
of Ref.~\cite{Calmet:2019eof}.
Similar calculations have been performed in e.g.~Refs.~\cite{Donoghue:2014yha,Codello:2015pga}.}
\begin{equation}	
	\ln\left(- \frac{\Box}{\mu^2} \right)\, f(x) = \sum_{k=0}^{\infty}\, c_{2k}\, r^{2k}
\end{equation}
with
\begin{equation}
	c_0 = - 2 \lim_{\epsilon\rightarrow 0} \left\{ 
	 \left[\gamma_{\rm E} - 1 + \ln(\mu\,\epsilon)  \right] f(0)
	+ \int_{\epsilon}^{\infty}  \frac{f(r)}{r} \, dr \right\}
	\ .
\end{equation}
Therefore, we find
\begin{align}
	\ln\left(- \frac{\Box}{\mu^2} \right) R &= d_0 + d_2\, r^2 + \mathcal{O}\left(r^{4}\right)
	\ ,
	\\
	\ln\left(- \frac{\Box}{\mu^2} \right) R^t_{\;t} &= x_0 + x_2\, r^2 + \mathcal{O}\left(r^{4}\right)
	\ ,
	\\
	\ln\left(- \frac{\Box}{\mu^2} \right) R^r_{\;r} &= y_0 + y_2\, r^2 + \mathcal{O}\left(r^{4}\right)
	\ ,
	\\
	\ln\left(- \frac{\Box}{\mu^2} \right) R^\theta_{\;\theta} &= z_0 + z_2\, r^2 + \mathcal{O}\left(r^{4}\right)
	\ .
\end{align}
These expressions lead to the corrections
\begin{align}
	\delta g_{tt}^{\rm NL} 
	&= - 16\, \pi\, \gn\, (1+a_0) \left[ 2\,\tilde{\alpha}\, d_2 
	-\tilde{\beta}\, \left(x_2 - y_2 - 2\, z_2 \right) \right]\,r^{2}
	+ \mathcal{O}\left(r^4\right)
	\ ,
	\\
	\delta g_{rr}^{\rm NL} 
	&= - 32\, \pi\, \gn \,\left\{ \tilde{\beta} \,(y_0 - z_0)
	+ \left[ 2 \,\tilde{\alpha}\, d_2 
	+ \tilde{\beta} \left( x_2 + 2\, y_2 - z_2 \right)
	\right]	r^2 \right\} 
	+ \mathcal{O}\left(r^4\right)
	\ .
\end{align}
Hence, if singularities are to be avoided, it is necessary to impose
\begin{equation}
	y_0=z_0
	\ .
\end{equation}
Interestingly, this condition can be translated into the following condition for the pressure anisotropy:~\footnote{Notice
that only smoothness, along with local time-independence and spherical symmetry are required.}
\begin{equation}\label{eq:AnisotropyCondition}
	\lim_{r\rightarrow 0} \ln\left(- \frac{\Box}{\mu^2} \right)
	\left( p_\perp - p_\parallel \right) = 0
	\ .
\end{equation}
If the time-independence and spherical symmetry holds globally, this is equivalent to
\begin{equation}\label{eq:AnisotropyCondition2}
	\int_{0}^{\infty} dr\, \frac{p_\perp(r) - p_\parallel(r)}{r} = 0
	\ .
\end{equation}
Moreover, if $p_\parallel$ is differentiable, conservation of the energy momentum tensor,
$\nabla^{\mu} T_{\mu\nu}=0$, implies that the above condition is equivalent to~\footnote{We assume $p_\parallel(\infty)=0$.}
\begin{equation}
	p_\parallel(0)
	=
	\int_0^\infty dr \, \frac{f'(r)}{2\,f(r)}
	\left[\rho(r) + p_\parallel(r) \right]
	\ .
\end{equation}
These identities are clearly satisfied for isotropic fluids, but cause trouble for many anisotropic fluids.
For instance, it can easily be verified that the condition is not satisfied for the Bardeen~\cite{Bardeen}
(see Appendix~\ref{A}), Hayward~\cite{Hayward:2005gi} and Frolov~\cite{Frolov:2014jva} metrics,
but it is satisfied for the Simpson-Visser metric~\cite{Simpson:2019mud} and for a constant density
star (cf. Ref.~\cite{Calmet:2019eof} for the explicit calculation.).
\section{Discussion}
\label{s:conc}
We have shown that perturbative quantum gravity can introduce singularities to geometries
that are classically regular.
Furthermore, for a locally spherically symmetric and time-independent space-time,
we have found a condition, given in Eq.~\eqref{eq:AnisotropyCondition}, on the pressure anisotropy
for which this scenario occurs.
Matter distributions that violate the condition contain a singularity at this order in perturbation theory.
\par                                                                                                                      
It should be stressed that the employed methods break down at and close to the emerging
singularity, as the truncation of the effective field theory becomes invalid.
Therefore it cannot be concluded that the singularity is physical.
In fact it is not unlikely that the newly found singularity is a spurious effect, and that it disappears,
when the full expansion of the effective action is taken into account and/or when the non-local
terms are evaluated using the complete curvature expansion.
Nevertheless such an effect is interesting, as it points out that for certain geometries naive
application of perturbation theory fails when calculating the quantum corrections at first order,
even when the classical energy density is close to the vacuum.
On the other hand, one cannot exclude the possibility that the singularity will persist,
when the effective action is considered up to infinite order.

\par
It would be interesting to investigate the higher order corrections within this framework. 
As the higher order corrections will come with different power of the Planck length, it is not expected that the terms at 2-loop or at any finite order will cancel the new singularities.~\footnote{We
might mention in passing the famous result by Goroff and Sagnotti that 2-loop corrections
diverge for pure gravity~\cite{Goroff:1985sz}.} 
However, such a higher order analysis could generate new singular terms and thus provide new constraints such as the one in
Eq.~\eqref{eq:AnisotropyCondition}, for which classical matter distributions are safe
in the sense that quantum corrections do not generate secular terms.
Moreover, a higher order analysis could provide indication whether resummation effects
can occur, which would indicate that the singularities are spurious, as they would be resolved at infinite order in perturbation theory.
\par
We will leave such an higher order analysis for future research.
Let us note however that higher order terms in the effective action are potentially
more dangerous than the first order terms analyzed in this paper.
Indeed, from dimensional analysis, one expects that local terms in the effective action
at order $\lp^{2k}$ generate corrections to the metric at order $\min\{r^{m-2k},r^{n-2k}\}$.
This was indeed found in Eqs.~\eqref{eq:LocCorTemp} and~\eqref{eq:LocCorRad}.
For $2k\geq \min\{m-2,n-2\}$, this generates terms that make the metric singular.
For $k=1$ we found that these dangerous terms are not present, as their coefficients vanish.
It is expected that the dangerous terms also vanish for $k>1$.                                                           
If such a cancellation mechanism were not present, this would pose new challenges for the use
of perturbative methods in quantum gravity.
Note that a priori there is no reason to expect that non-perturbative quantum gravity
should be invoked, as the classical densities and pressures we consider here are well below
the Planck scale.
For the same reason, we do not expect that the singularity we found can be removed 
by any matter rearrangements which keep density and pressure in the sub-Planckian range.
\par
Furthermore, following the same reasoning, we find that, if $2<m,n<\infty$, the quantum corrections
will always generate corrections with smaller powers of $r$.
In particular, the non-local terms are expected to generate corrections at order $r^2$.
Therefore, a byproduct of our analysis is that a regular and smooth quantum space-time,
that is locally spherically symmetric and time-independent should always
have the form of Eq.~\eqref{eq:metric} with $f$ and $g$ given by Eqs.~\eqref{eq:smoothmetricf}
and~\eqref{eq:smoothmetricg}, with the extra assumption that all coefficients $a_{2n}$ and $b_{2n}$
are non-zero unless some kind of fine-tuning occurs.
In addition, $a_0>-1$, and using the analysis in section~\ref{s:ec}, one can impose $b_2<0$,
and $|a_2| \leq (1+a_0)|b_2|$.
\par
Finally, one could try to generalize these results to space-times that do not have
local spherical symmetry or are time-dependent, as it could lead to similar conditions
on the expansion of the metric components.
We will leave this for a future paper.
\section*{Acknowledgments}
The work of X.C.~is supported in part  by the Science and Technology Facilities Council (grant number ST/P000819/1).
R.C.~is partially supported by the INFN grant FLAG and his work has also been carried out in the framework of activities
of the National Group of Mathematical Physics (GNFM, INdAM) and COST action {\em Cantata\/}. 
The work of F.K.~is supported by a doctoral studentship of the Science and Technology Facilities Council.
F.K.~is grateful for the hospitality of the Universit\`a di Bologna, where most of this work was carried out.
\appendix
\section{The Bardeen Metric}
\label{A}
Let us consider the Bardeen metric~\cite{Bardeen} as an example of regular space-time where one encounters
singularities of the type discussed in this work.
The Bardeen space-time has no central singularity but a de~Sitter core (see Eq.~\eqref{eq:dscore}),
and can be motivated by coupling Einstein gravity to a non-linear electrodynamic field~\cite{AyonBeato:2000zs}.
The Bardeen line element is of the form in Eq.~\eqref{eq:metric} with
\begin{equation}
	f(r) = g(r) =
	1 - \frac{2 \,\gn\, M\, r^2}{\left(r^2+l^2\right)^{3/2}}
	\ ,
\end{equation}
where $l>0$ is some length scale.
For sufficiently small values of $l$, this space-time contains a horizon and, in fact, it reduces to the Schwarzschild
geometry in the limit $l\to 0$ and to the Minkowski space-time in the limit $l\to\infty$.
\par
Using the procedure outlined in section~\ref{s:qc}, one can calculate the metric corrections.
Up to $\mathcal{O}(\gn^3)$ the local corrections are given by
\begin{align}
	\delta g_{tt}^{\rm L} 
	&= 
	- \frac{192\, \pi\, \hbar \,\gn^2\, M\, l^2}{\left(r^2 + l^2\right)^{7/2}}
	\left[\tilde{c}_1(\mu) \left(  r^2 - 4\, l^2\right)
	-  \tilde{c}_2(\mu) \left(  r^2 + l^2\right)\right],\\
	\delta g_{rr}^{\rm L} 
	&= 
	\frac{480\, \pi\, \hbar\, \gn^2\, M \, l^2\, r^2}{\left(r^2 + l^2\right)^{9/2}}
	\left[2\, \tilde{c}_1(\mu) \left(  r^2 - 6\, l^2\right)
	+  \tilde{c}_2(\mu) \left( 2\, r^2 - 5\, l^2\right)\right],
\end{align}
and the non-local corrections are given by
\begin{align}
	\delta g_{tt}^{\rm NL} 
	&= 
	\frac{128\, \pi\, \hbar\, \gn^2\, M}{\left(r^2 + l^2\right)^{7/2}}
	\left\{ \tilde{\alpha} \left[  r^4  + 16\, l^2\, r^2 - 31\,l^4 
	- 3\,l^2\left(r^2 - 4\, l^2 \right) \left(\gamma_{\rm E} + \ln\left[\frac{2\,\mu\left(r^2 + l^2 \right)}{l} \right] \right) \right]
	\right. \nonumber\\
	&\qquad \qquad \quad \left.
	+ \tilde{\beta} \left[  r^4  - 6\, l^2\, r^2 - 7\, l^4 
	+ 3\,l^2 \left(r^2 + l^2 \right) \left(\gamma_{\rm E} + \ln\left[\frac{2\,\mu\left(r^2 + l^2 \right)}{l} \right] \right) \right]\right\},\\
	\delta g_{rr}^{\rm NL} 
	&= 
	- \frac{64\, \pi\, \hbar\, \gn^2\, M}{\left(r^2 + l^2\right)^{9/2}}
	\left\{ 2\, \tilde{\alpha}\, r^2 \left[ 3 \, r^4  + 82\, l^2\, r^2 - 273\, l^4 
	- 15\,l^2\left(r^2 - 6\, l^2 \right) \left(\gamma_{\rm E} + \ln\left[\frac{2\,\mu\left(r^2 + l^2 \right)}{l} \right] \right) \right]
	\right. \nonumber\\
	&\quad  \left.
	+ \tilde{\beta}\, l^2 \left[ 125\, r^4  - 224\, l^2\, r^2 + 3\, l^4 
	- 15\, r^2 \left(2\, r^2 - 5\, l^2 \right) \left(\gamma_{\rm E} + \ln\left[\frac{2\,\mu\left(r^2 + l^2 \right)}{l} \right] \right) \right]\right\}
	\ ,
\end{align}
where $\gamma_{\rm E}$ is Euler constant.
Using these expressions, one can calculate the quantum corrected Ricci scalar, which we split
into the classical Ricci scalar and a quantum correction as
\begin{equation}
	R = R^{\rm c} + R^{\rm q}_{\rm fin} +  R^{\rm q}_{\rm div}
	\ .
\end{equation}
The classical part $R^{\rm c}$ is finite everywhere, while the quantum correction
contains both a finite contribution $R^{\rm q}_{\rm fin}$ and a contribution
\begin{equation}
	R^{\rm q}_{\rm div} = - \frac{384 \, \pi \, \tilde{\beta} \, \hbar \, \gn^2 \, M \, l^8}{r^2 \left( r^2 + l^2 \right)^{11/2}}
\end{equation}
which diverges for $r\to 0$.
This divergence cannot be canceled within perturbation theory, since corrections only appear at
$\mathcal{O}(\gn^3)$.
Resolution of this singularity can thus only be achieved in a non-perturbative way.
Notice, however, that this singularity is integrable in the sense that radial geodesics can be extended through the singularity.
Furthermore, the expansion in $\gn$ consists of both a classical expansion in $\gn M/l$
and a quantum expansion in $\hbar\, \gn/l^2$.
To keep the classical expansion under control we will thus assume $l> \gn M$.
For this choice the space-time does not contain a horizon, implying that the singularity is naked.


\begin{thebibliography}{99}
	\bibitem{Penrose:1964wq} 
	R.~Penrose,
	``Gravitational collapse and space-time singularities,''
	Phys.\ Rev.\ Lett.\  {\bf 14}, 57 (1965).
	
	
	\bibitem{Hawking:1966vg} 
	S.~W.~Hawking,
	``Singularities in the universe,''
	Phys.\ Rev.\ Lett.\  {\bf 17}, 444 (1966).
	
	
	\bibitem{Geroch:1987qn} 
	R.~P.~Geroch and J.~H.~Traschen,
	``Strings and Other Distributional Sources in General Relativity,''
	Phys.\ Rev.\ D {\bf 36}, 1017 (1987)
	[Conf.\ Proc.\ C {\bf 861214}, 138 (1986)].
	
	
	\bibitem{Weinberg:1980gg} 
	S.~Weinberg,
	``Ultraviolet Divergences In Quantum Theories Of Gravitation,''
	in \textit{General Relativity: An Einstein Centenery Survey}, Cambridge, UK, 790 (1980).
	
	
	\bibitem{Barvinsky:1984jd} 
	A.~O.~Barvinsky and G.~A.~Vilkovisky,
	``The Generalized Schwinger-de Witt Technique And The Unique Effective Action In Quantum Gravity,''
	Phys.\ Lett.\  {\bf 131B}, 313 (1983).
	
	
	\bibitem{Barvinsky:1985an} 
	A.~O.~Barvinsky and G.~A.~Vilkovisky,
	``The Generalized Schwinger-Dewitt Technique in Gauge Theories and Quantum Gravity,''
	Phys.\ Rept.\  {\bf 119}, 1 (1985).
	
	
	\bibitem{Barvinsky:1987uw} 
	A.~O.~Barvinsky and G.~A.~Vilkovisky,
	``Beyond the Schwinger-Dewitt Technique: Converting Loops Into Trees and In-In Currents,''
	Nucl.\ Phys.\ B {\bf 282}, 163 (1987).
	
	
	\bibitem{Barvinsky:1990up} 
	A.~O.~Barvinsky and G.~A.~Vilkovisky,
	``Covariant perturbation theory. 2: Second order in the curvature. General algorithms,''
	Nucl.\ Phys.\ B {\bf 333}, 471 (1990).
	
	
	\bibitem{Buchbinder:1992rb} 
	I.~L.~Buchbinder, S.~D.~Odintsov and I.~L.~Shapiro,
	``Effective action in quantum gravity,''
	Bristol, UK: IOP (1992) 413 p.
	
	
	\bibitem{Donoghue:1994dn} 
	J.~F.~Donoghue,
	``General relativity as an effective field theory: The leading quantum corrections,''
	Phys.\ Rev.\ D {\bf 50}, 3874 (1994).
	
	
	\bibitem{Stelle:1977ry} 
	K.~S.~Stelle,
	``Classical Gravity with Higher Derivatives,''
	Gen.\ Rel.\ Grav.\  {\bf 9}, 353 (1978).
	
	
	\bibitem{Donoghue:1993eb} 
	J.~F.~Donoghue,
	``Leading quantum correction to the Newtonian potential,''
	Phys.\ Rev.\ Lett.\  {\bf 72}, 2996 (1994).
	
	
	\bibitem{Calmet:2018qwg} 
	X.~Calmet and B.~Latosh,
	``Three Waves for Quantum Gravity,''
	Eur.\ Phys.\ J.\ C {\bf 78}, no. 3, 205 (2018).
	
	
	\bibitem{Kuipers:2019qby} 
	F.~Kuipers and X.~Calmet,
	``Singularity theorems in the effective field theory for quantum gravity at second order in curvature,''
	arXiv:1911.05571 [gr-qc].
	
	
	\bibitem{Kuntz:2019lzq} 
	I.~Kuntz and R.~Casadio,
	``Singularity avoidance in quantum gravity,''
	Phys.\ Lett.\ B {\bf 802}, 135219 (2020).
	
	\bibitem{Frolov:1981qr}
	V.~P.~Frolov and G.~Vilkovisky,
	``Spherically Symmetric Collapse in Quantum Gravity,''
	Phys.\ Lett.\ B {\bf 106}, no. 4, 307 (1981)
	
	\bibitem{Donoghue:2014yha} 
	J.~F.~Donoghue and B.~K.~El-Menoufi,
	``Nonlocal quantum effects in cosmology: Quantum memory, nonlocal FLRW equations, and singularity avoidance,''
	Phys.\ Rev.\ D {\bf 89}, no. 10, 104062 (2014).
	
	\bibitem{Calmet:2017qqa}
	X.~Calmet and B.~K.~El-Menoufi,
	``Quantum Corrections to Schwarzschild Black Hole,''
	Eur. Phys. J. C \textbf{77}, no.4, 243 (2017)
	
	\bibitem{Calmet:2019eof} 
	X.~Calmet, R.~Casadio and F.~Kuipers,
	``Quantum Gravitational Corrections to a Star Metric and the Black Hole Limit,''
	Phys.\ Rev.\ D {\bf 100}, no. 8, 086010 (2019).
	
	\bibitem{Deser:1974cz} 
	S.~Deser and P.~van Nieuwenhuizen,
	``One Loop Divergences of Quantized Einstein-Maxwell Fields,''
	Phys.\ Rev.\ D {\bf 10}, 401 (1974).
	
	
	\bibitem{Birrell:1982ix} 
	N.~D.~Birrell and P.~C.~W.~Davies,
	``Quantum Fields in Curved Space,''
	
	
	\bibitem{Kallosh:1978wt} 
	R.~E.~Kallosh, O.~V.~Tarasov and I.~V.~Tyutin,
	``One Loop Finiteness Of Quantum Gravity Off Mass Shell,''
	Nucl.\ Phys.\ B {\bf 137}, 145 (1978).
	
	
	\bibitem{Vilkovisky:1984st} 
	G.~A.~Vilkovisky,
	``The Unique Effective Action in Quantum Field Theory,''
	Nucl.\ Phys.\ B {\bf 234}, 125 (1984).
	
	\bibitem{Codello:2015pga}
	A.~Codello and R.~K.~Jain,
	``On the covariant formalism of the effective field theory of gravity and its cosmological implications,''
	Class. Quant. Grav. \textbf{34}, no.3, 035015 (2017)
	
	\bibitem{Bardeen}
	J.~M.~Bardeen
	``Non-singular general-relativistic gravitational collapse''
	in \textit{Proceedings of International Conference GR5}, Tbilisi, Georgia, 174 (1968).
	
	\bibitem{Hayward:2005gi} 
	S.~A.~Hayward,
	``Formation and evaporation of regular black holes,''
	Phys.\ Rev.\ Lett.\  {\bf 96}, 031103 (2006).
	
	
	\bibitem{Frolov:2014jva} 
	V.~P.~Frolov,
	``Information loss problem and a 'black hole` model with a closed apparent horizon,''
	JHEP {\bf 1405}, 049 (2014).
	
	
	\bibitem{Simpson:2019mud} 
	A.~Simpson and M.~Visser,
	``Regular black holes with asymptotically Minkowski cores,''
	Universe {\bf 6}, no. 1, 8 (2019).
	
	\bibitem{AyonBeato:2000zs}
	E.~Ayon-Beato and A.~Garcia,
	``The Bardeen model as a nonlinear magnetic monopole,''
	Phys. Lett. B \textbf{493}, 149-152 (2000)
	
	\bibitem{Goroff:1985sz}
	M.~H.~Goroff and A.~Sagnotti,
	``Quantum Gravity at Two Loops,''
	Phys. Lett. B \textbf{160}, 81-86 (1985).
	
	
\end{thebibliography}
\end{document}